\documentclass[12pt,preprint]{aastex}

\newcommand{\ee}[1]{\mbox{${} \times 10^{#1}$}}

\newcommand{\kms}{\mbox{km s$^{-1}$}}

\newcommand{\sfr }{\mbox{$\dot M_{\star}$}}

\newcommand{\lsun}{\mbox{L$_\odot$}}
\newcommand{\msun}{\mbox{M$_\odot$}}
\newcommand{\ta}{\mbox{$T_A^*$}}

\newcommand{\tr}{\mbox{$T_R$}}

\newcommand{\mean}[1]{\mbox{$\langle#1\rangle$}} 

\newcommand{\jj}[2]{\mbox{$J = #1\rightarrow#2$}}

\newcommand{\lfir}{\mbox{$L_{IR}$}}
\newcommand{\lmol}{\mbox{$L_{mol}$}}
\newcommand{\lhcn}{\mbox{$L_{HCN}$}}
\newcommand{\lco}{\mbox{$L_{CO}$}}
\newcommand{\lunit}{\mbox{$L_{unit}$}}

\newcommand{\mstar}{\mbox{$M_{\star}$}}
\newcommand{\lstar }{\mbox{$L_{\star}$}}

\newcommand{\mdense}{\mbox{$M(dense)$}}

\shorttitle{Inflow in Massive Regions}
\shortauthors{Wu \& Evans }
\begin{document}
\title {\bf Connecting Dense Gas Tracers of Star Formation in our Galaxy to 
High-z Star Formation }
\author {Jingwen Wu$^{1}$, Neal J. Evans II$^{1}$, Yu Gao$^{2}$, 
 Philip M. Solomon$^{3}$, Yancy L. Shirley$^{4,5}$, Paul A. Vanden Bout$^{6}$ }
\altaffiltext{1}{Department of Astronomy, The University of Texas at Austin,
        1 University Station, C1400, Austin, Texas 78712--0259,
        jingwen@astro.as.utexas.edu, nje@astro.as.utexas.edu}

\altaffiltext{2}{Purple Mountain Observatory, Chinese Academy of Sciences, 
2 West Beijing Road, Nanjing 210008. P.R.China, yugao@pmo.ac.cn}

\altaffiltext{3}{Department of Physics and Astronomy, SUNY at Stony Brook, Stony Brook, NY 11974, Philip.Solomon@sunysb.edu}

\altaffiltext{4}{Steward Observatory, University of Arizona, Tucson, AZ 85721.
 yshirley@as.arizona.edu}
\altaffiltext{5}{Bart J. Bok Fellow.}

\altaffiltext{6}{National Radio Astronomy Observatory, 520 Edgemont Rd., 
Charlottesville, VA 22903. pvandenb@nrao.edu}


\pagebreak

\begin{abstract}

Observations have revealed prodigious amounts of star formation in starburst 
galaxies as traced by dust and molecular emission, even at large redshifts. 
Recent work shows that for both nearby spiral galaxies and 
distant starbursts, the global star formation rate, as indicated by the 
infrared 
luminosity, has a tight and almost linear correlation with the amount of 
dense gas as traced by the luminosity of HCN. Our surveys of Galactic dense 
cores in HCN 1$-$0 emission show that this correlation continues 
to a much smaller scale, with nearly the same ratio of infrared
luminosity to HCN luminosity found over 7-8 orders of magnitude in \lfir, 
with a lower cutoff around $10^{4.5}$ \lsun\ of infrared luminosity.
The linear correlation suggests that we may understand distant star 
formation in terms of the known properties of local star-forming 
regions. Both the correlation and the luminosity cutoff
can be explained if the basic unit of star formation in galaxies is
a dense core, similar to those studied in our Galaxy.

\end{abstract}

\clearpage

\section{Introduction}

Recent work has revealed large amounts of dust and molecules in 
starburst galaxies, even at large redshifts (e.g. Isaak et al. 2002; 
Reuland et al. 2003; Greve et al. 2005; Solomon \& Vanden Bout 2005). 
Understanding star formation in galaxies at high redshift is a critical step
in understanding the formation of galaxies in the early Universe.

The simplest and most widely used relations between the star formation rate 
and a property of the interstellar medium (ISM) are the so-called
``Schmidt laws". Schmidt (1959) proposed that the star formation rate was
proportional to $\rho^2$, where $\rho$ is the gas volume density. 
In their modern form, these ``laws" relate the {\it surface density} of
star formation to the {\it surface density} of gas:
\begin{equation}
 \Sigma_{SFR} = A \Sigma^{N}_{gas} 
\end{equation}
(e.g., Kennicutt 1998).
The index $N$ has been inferred by various authors to be in the range of
1 to 2 (Kennicutt 1997). Measurements of HI, CO, and H$\alpha$ on a large 
sample of normal spiral galaxies and starburst galaxies (Kennicutt 1998) 
could be fitted  with $N = 1.4$. It should be noted that for many of these 
galaxies, particularly the starburst galaxies, in the Kennicutt sample the 
surface density or surface brightness was not measured but obtained only by 
dividing the luminosity by a characteristic size often obtained from another 
parameter. 

On a global scale, including luminous and ultraluminous starburst galaxies,
there is a correlation between the total luminosity of far-infrared emission,
which traces the star formation rate (e.g. Sanders and Mirabel 1996; Kewley 
et al. 2002),
and the total luminosity of CO, tracing the molecular gas mass. 
However, this relation is not linear; the ratio of \lfir\ to \lco\ increases
with increasing \lfir\  ( Sanders and Mirabel 1996; Solomon et al. 1997; 
Gao \& Solomon 2004a, 2004b; Solomon \& Vanden Bout 2005).
Does this variation reflect an increase in ``efficiency" of star formation,
or an increasing failure of CO to trace the gas that is relevant to star 
formation?

The latter possibility is suggested by data on HCN $J = 1-0$ in galaxies.
A recent survey of HCN $J=1-0$ emission in 65 
normal spiral and starburst galaxies found that the star formation rate, 
as traced by the infrared luminosity, has a tight and linear correlation 
with the luminosity of HCN (Gao \& Solomon 2004a, 2004b). 
Those authors argued that CO is not linearly correlated with star formation 
because it traces only the low density GMC envelopes, not the really active 
star-forming part, the dense cores. 
The critical molecular parameter that measures 
star formation rates in galaxies is the amount of dense molecular gas,
measured by the HCN luminosity.  Because
HCN traces the dense gas better than CO,  L$_{HCN}$ has a tighter correlation 
with 
$L_{IR}$ than does $L_{CO}$.  The correlation remains linear over a factor 
of $10^3$ in luminosity for both normal galaxies and extreme starbursts,
like luminous and ultraluminous infrared galaxies (LIRGs and ULIRGs, see 
Sanders and Mirabel, 1996). Gao and Solomon (2004a) therefore argue that both 
normal galaxies and starbursts should have the same star formation rate per 
amount of {\it dense} gas. 

At lower luminosities, the relation for galaxies between \lfir\ and \lco\ is 
linear, but the ratio between \lfir\ and \lco\ has a dispersion of 
an order of magnitude. 
An even larger variation in \lfir/\lco\ (several orders of magnitude) is seen 
 in Galactic clouds (Mooney \& Solomon 1988; Evans 1991; Mead et al. 1990).
The dispersion in the relation for galaxies is less if \lhcn\ is used instead 
of CO; the same is true for other tracers of dense gas in Galactic cores 
(e.g. Mueller et al. 2002, Shirley et al. 2003), 
suggesting that studies
of Galactic star formation can shed light on the trends in other galaxies. 

In the Milky Way, star formation is dominated by 
clustered star formation (Carpenter 2000; Lada \& Lada 2003 ). 
Clustered star formation produces stars with a range of masses, 
but massive stars form nearly exclusively in clusters within massive 
dense cores. Since massive stars dominate the luminosity, they 
are the stars directly probed in studying star formation in other galaxies. 
Thus, understanding the relation between star formation and massive,
dense cores in our Galaxy may shed light on the starburst phenomenon.

These cores are dense, turbulent, and dusty. They are well identified by 
the continuum emission from dust (e.g., Mueller et al. 2002) and line emission 
from molecular dense gas tracers like CS (Plume et al. 1992, 1997; Shirley 
et al. 2003) and HCN (Wu \& Evans 2003). In our previous work on a large 
sample of massive cores, we found that the bolometric (almost all far-infrared)
luminosity of the cores is roughly proportional to the mass inferred from the
dust emission (Mueller et al. 2002) and the virial mass determined from CS
(Shirley et al. 2003). This result suggests that a relation between
\lfir\ and \lhcn\ may exist in Galactic dense cores, possibly similar to that 
in starburst galaxies.

One difficulty is that systematic studies of the Galactic dense cores have used
dust continuum emission and CS, while HCN has been more commonly used for 
studies of
galaxies. To facilitate comparison with the HCN $J = 1-0$ galaxy survey, 
we have mapped the HCN $J = 1-0$ transition in a sample of 
47 Galactic star forming cores (Wu et al. 2005, in preparation). 
In this paper we will 
summarize the results from this survey in connection with the HCN surveys of
other galaxies.

The sample mapped in HCN $J = 1-0$ is mostly a subset of a larger sample of 
Galactic massive dense cores that have been mapped in CS and dust emission
(Shirley et al. 2003; Mueller et al. 2002). The sources in this category 
have infrared luminosities ranging from 10$^{3}$ - 10$^{7}$ L$_{\odot}$ 
and most contain compact or ultracompact H $\amalg$ (UCH$\amalg$) regions.
To extend the sample towards lower luminosities, 
we selected 14 IRAS sources from outflow surveys (Zhang et al. 2005; Wu et al. 
2004) and a few from other publications. 

\section{Observation \& Data Analysis}

Observations of HCN \jj10\ (88.6318473 GHz) on Galactic dense cores 
were made with the 
14-m telescope of the Five College Radio Astronomy Observatory (FCRAO) 
in 2004 April, December and 2005 February. The FWHM of the beam for 
this frequency is 58\arcsec. 
The 16-element focal plane array (SEQUOIA) was used, with typical system 
temperatures 100-200K. A velocity resolution of 0.1 \kms\ was achieved with 
the 25 MHz bandwidth on the dual channel correlator (DCC). 
We convert the measured \ta\ to \tr\ via $\tr = \ta/(\eta_{FSS}\eta_c)$, with
$\eta_{FSS}=0.7$. The value of $\eta_{c}$ depends on source size; for
the typical map in this study ($\sim 10\arcmin$), $\eta_{C}=0.7$.
The map size was extended until the edge of the HCN \jj10\ emission was 
reached, typically at the $2\sigma$ level 
(mean $\sigma$ $\sim$ 0.3 K km/s), so we could get the total 
HCN luminosity.

Maps of HCN 1-0 emission were obtained for 47 Galactic star-forming cores. 
Detailed results will be presented elsewhere (Wu et al. 2005, in preparation),
but we give a few properties of the sample here, 
which are relevant to this paper. 
More than 90\% of the cores were well resolved by the maps. We focus on 
these resolved sources in this paper. 

The size of the core is characterized by the nominal core radius after
beam deconvolution, 
$R_{HCN}$, the radius of a circle that has the same area as the half 
peak intensity contour:
$$R_{HCN}=D(\frac{A_{1/2}}{\pi}-\frac{\theta^{2}_{beam}}{4})^{1/2}, $$ 
where $A_{1/2}$ is the area within the contour of half peak intensity.
The median $R_{HCN 1-0}$ of the sample is 0.48 pc. 

The HCN line luminosity of each core, assuming a gaussian 
brightness distribution for the source and a gaussian beam, is: 
\begin{equation}
L_{HCN}=23.5\times 10^{-6}\times D^{2}\times (\frac{\pi\times\theta_{s}^{2}}
{4ln 2})\times(\frac{\theta_{s}^{2}+\theta_{beam}^{2}}{\theta_{s}^{2}})
\times \int T_{R}dv 
\end{equation}
Here D is the distance in the unit of kpc, $\theta_{s}$ and $\theta_{beam}$ 
are the angular size of the source and beam in arcsecond. 
This method is parallel to that of Gao \& Solomon (2004b), but adapted
to Galactic cores. \lhcn\ ranges from 0.4 to 8000 
K \kms\ pc$^{2}$, with the median value of 80 K \kms\ pc$^{2}$ . 

The total infrared luminosity (8-1000 $\mu$m) was calculated 
based on the 4 IRAS bands (Sanders and Mirabel 1996), as was done for
the galaxy sample of Gao \& Solomon (2004a):
\begin{equation}
 \lfir = 0.56\times D^{2}\times (13.48\times f_{12}+5.16\times f_{25}+
2.58\times f_{60}+f_{100}), 
\end{equation}
where $f_x$ is the flux in band $x$ from the four IRAS bands in the units 
of Jy, D in kpc, and L$_{IR}$ in \lsun.

\section{Comparison of Milky Way and galactic Relations}

The derived \lfir\ and \lhcn\ are plotted in a log-log diagram in 
Fig. 1 to compare with data on galaxies from Gao \& Solomon (2004a). 
The correlation of \lfir--\lhcn\ 
extends from galaxy scales to the much smaller scales of Galactic molecular 
cores. The fit for Gao \& Solomon's 
galaxy sample is log$(\lfir)=1.00\times {\rm log}(\lhcn)+2.9$,
 without a few galaxies that only have upper limits to HCN emission. 
This linear correlation continues to the Galactic massive cores, but a decline
in \lfir\ occurs at $10^{4.5}$ \lsun,
below which the slope of the correlation becomes much steeper. 
When fit to Galactic cores with $\lfir > 10^{4.5}$ \lsun, the linear 
least squares fit gives 
log($\lfir)=1.02(\pm0.06)\times {\rm log}(\lhcn)+2.79(\pm0.16)$. 
This relation agrees remarkably well with the relation for galaxies, as seen 
in fig. 1a. In fig. 1b, we fit simultaneously the data of galaxies from 
Gao \& Solomon (2004a) and Galactic cores with $\lfir > 10^{4.5}$ \lsun. 
The result is log$(\lfir)=1.01\times {\rm log}(\lhcn)+2.83$, with 
a correlation coefficient of 0.99.
Understanding the physics behind this linear correlation will lead to
a better understanding of star formation on galactic scales. 

The linear log($\lfir)-{\rm log}(\lhcn)$ 
correlation and the turnoff can be seen more 
clearly from fig. 2, where the distance independent ratio 
\lfir/\lhcn\ has been plotted versus \lfir\ (top) and against \lmol\ (bottom).
A constant mean value of \lfir/\lhcn\ is seen over 8 orders of magnitude
in \lfir, from galaxies to Galactic cores, as long as $\lfir > L(min)$,
with $L(min)$ being around $10^{4.5}$ \lsun. The corresponding cut-off
value for \lhcn\ is $\lunit = 10^{1.5}$ K \kms\ pc$^{2}$.
For Galactic cores,
$\mean{\lfir/\lhcn}= 911\pm 227$, with median 380. For galaxies, 
$\mean{\lfir/\lhcn}= 950\pm 76$, with median 855. The uncertainties are
the standard deviation of the means, which are remarkably similar, though
the dispersion is much higher for Galactic cores. The median 
for dense cores is significantly smaller than its mean, which indicates that 
the mean is dominated by a few quite large values. The logarithmic mean and 
median are 2.64$\pm$0.53 and 2.54$\pm$0.56 for dense cores with 
$\lhcn > \lunit$, and 2.91$\pm$0.24 and 2.93$\pm$0.25 for galaxies.

For comparison, we added CO data on Galactic cores (Mooney \& Solomon 1988), 
galaxies (Gao \& Solomon 2004a) and high-z molecular emission line galaxies 
(Solomon \& Vanden Bout 2005) in fig. 2.
The ratio, \lfir/\lco, changes by two orders of magnitude as \lco\ increases
from Galactic cores to distant galaxies,
confirming Gao \& Solomon's conclusion that CO is 
not as good a tracer of star-forming gas as is HCN, especially for very 
luminous starburst galaxies.

\section{Discussion}

The fact that \lfir/\lhcn\ is similar, on average, for star-forming
cores in the Galaxy, normal spirals, starbursts, and ULIRGs
suggests the possibility of interpreting intense high-z star formation in
terms of nearby high mass star forming regions. 
Before we can exploit this possibility, we must understand some key points.
Why does \lfir/\lhcn\ rise steeply with \lhcn\ and
then remain constant for $\lhcn > \lunit$? 
And why is the ratio, \lfir/\lco, NOT constant for starbursts?

As a first step, we seek a more physical basis for the relations.
We have so far discussed \lhcn\ as a measure of the mass of dense gas, but
can we quantify this assumption? A roughly linear correlation between the 
mass of dense gas and bolometric luminosity has been found by 
our work on CS and dust emission (Shirley et al. 2003; Mueller et al. 2002). 
To see whether this applies also to HCN \jj10,
we have calculated the virial mass of the dense gas
(\mdense) using the most optically thin line (C$^{34}$S \jj54) to measure the 
linewidth 
and compared \mdense\ to \lhcn. Based on the 31 cores with available
C$^{34}$S \jj54 data, we obtained the correlation: 
log($M_{vir}(R_{HCN1-0})$)=(0.81$\pm$0.03)$\times$ log($L_{HCN1-0}$)
+1.29($\pm$0.09). The correlation is roughly linear. The 
$M_{vir}$-$L_{HCN1-0}$ plot and details of the observations 
of C$^{34}$S \jj54 will be presented by Wu et al. (2005, in preparation).
Leaving out one peculiar source (G35.58-0.03), we get $\mean{\mdense/\lhcn} = 
11\pm 2$~\msun/K \kms\ pc$^{2}$, where the uncertainty is the standard 
deviation of the mean; the median value is 6 ~\msun/K \kms\ pc$^{2}$, 
indicating that the mean is affected by some quite large
values. The logarithmic mean is 7$\pm$2 ~\msun/K \kms\ pc$^{2}$. 
Some of the scatter in the ratio may be caused by distance 
uncertainties because the virial mass depends linearly on distance, while
$\lhcn \propto D^2$.

Even after establishing that \lhcn\ traces the mass of dense gas,
it is not at all clear why \lfir/\lhcn\ should be constant, since 
the luminosity of a cluster is typically dominated by its most massive 
members, and $\lstar \propto \mstar^\alpha$, with $\alpha \approx 3$ to 4.
Indeed, below the cutoff of $10^{4.5}$ \lsun, \lfir\ does rise rapidly
with \mdense. What causes the transition to a constant value?

To solve this puzzle, we propose the existence of a basic unit of cluster 
formation. For
\mdense\ less than the mass of this unit, \lfir/\mdense\ rises rapidly
with \mdense, as higher mass stars can form. For \mdense\ greater than the
mass of this unit, the IMF is reasonably sampled and further increases
in mass produce more units, but no further change in \lfir/\mdense.
If we suppose that larger scale cluster formation is built up 
by adding more and more such units,
then the linear correlation between the total \lfir\ and 
\mdense\ is a natural result. In that case, the only difference between star 
formation on different scales and in different environments--big clusters, 
normal galaxies, massive ULIRGs--is just how {\it many} such cores 
they contain. This is of course somewhat of a simplification because
the cores have a range of properties. The fact that the mean is roughly
twice the median for both \lfir/\lhcn\ and \mdense/\lhcn\ for Galactic cores
suggests that the most extreme cases may dominate when averaged over a whole
galaxy.  Since most of our cores contain compact H $\amalg$ or 
UCH $\amalg$ regions, which trace the most massive stars, it is interesting
to compare to the luminosity function of these regions.
A study of Galactic UCH $\amalg$ regions 
did find a peak luminosity ($\sim$10$^{5}$\lsun ) that is close to  that of
a basic unit (Cassassus et al. 2000). 
The detailed discussion of this model will be presented in a 
separate paper (Wu et al. 2005, in preparation).

Another question is why does \lfir/\lco\ increase as we move from Galactic 
Cores or normal spiral galaxies to starburst galaxies. 
In Galactic clouds, CO can be used as a tracer of the overall
mass of molecular clouds, even though it is optically thick and thermalized.
However, CO does not specifically trace the mass of dense cores. For that, dust
continuum emission, or
molecules, like HCN, that are only excited at higher densities are required.
In the Galaxy, these dense cores occupy a small fraction of the mass
of a cloud (typically a  few percent).
If the fraction of dense gas in the overall cloud
stays roughly constant, the relation between \lfir\ and \lco\ can stay 
linear, but this relation is secondary to that between \lfir\ and \mdense.  
For starburst galaxies, the fraction of the molecular gas concentrated in 
dense cores increases, causing (Gao \& Solomon 2004a) the secondary relation 
between \lfir\ and \lco\ to become non-linear.

These considerations lead us to offer some new versions of ``Schmidt
Laws":
\begin{equation}
\sfr (\msun yr^{-1}) \sim 1.4\ee{-7} \lhcn ( K \kms\ pc^2)
\end{equation}
\begin{equation}
\sfr (\msun yr^{-1}) \sim 1.2\ee{-8} \mdense (\msun)
\end{equation}
where \sfr\ is the star formation rate, and we have assumed that 
$\sfr (\msun yr^{-1}) =  2.0\ee{-10} \lfir (\lsun)$
(see Gao \& Solomon (2004a)) and we use the fit to both dense cores
and galaxies. The coefficients (1.4\ee{-7} and 1.2\ee{-8}) are very 
similar to, but slightly less than, those values (1.8\ee{-7} and 1.8\ee{-8}) 
given by Gao \& Solomon (2004a), based only on the galaxies.

Given these simpler, linear
relations, how do we understand the usual star formation law (Kennicutt 1998) 
relations, with a steeper dependence 
($\Sigma_{SFR} \propto \Sigma^{1.4}_{gas}$) of \sfr\ on gas mass? 
We suggest that the 
steeper dependence reflects the dependence of \mdense\ on the
surface density, or total mass, of gas. All the tracers of gas
used by Kennicutt trace lower density components, not the actual
gas that is directly involved in star formation. Once it is clear that it is 
the dense gas mass that indicates the star formation rate, it becomes clear 
why the total surface density of gas may not be a clean star formation 
indicator. For example, there is no evidence that HI emission in galaxies 
correlates at all with the star formation rate deduced form the far-infrared 
luminosity, so its contribution to the surface density may have no effect on 
the star formation rate.  Attempts to provide
a theoretical framework (e.g., Li et al. 2005, Krumholz \& McKee 2005)
for the Kennicutt relations should also be
able to explain the new relations.

It will be interesting to see how far these ideas can be extended.
One future project is to look at HCN in more nearby galaxies, 
especially in individual regions forming super-star clusters, which 
may be the building blocks of star formation in starburst regions 
like ULIRGs and LIRGs (Ho 1997).
ALMA will allow detailed study of HCN in other galaxies, including
the \jj32\ transition.
Another is to explore how far toward higher z these relations
can be pushed, and to understand how the relations depend on
metallicity and chemistry. Theoretical work by Lintott et al. (2005) suggest
that HCN may trace very early star formation, depending on the
nucleosynthetic yield of the earliest stars.

\acknowledgements

We are grateful to Mark Heyer and other FCRAO staff for assistance with the 
observations. We thank the referee for helpful comments. 
We thank Dan Jaffe, John Scalo, Shardha Jogee, John Kormendy, Michael Dopita,
and Luis Ho for helpful discussions. 
This work was supported by NSF Grant AST-0307250 to the
University of Texas at Austin and by the State of Texas. YG acknowledges
support of the NSF of China grants 10425313 \& 10333060.

\clearpage


\begin{figure}[hbt!]
\rotatebox{-90}{\plotone{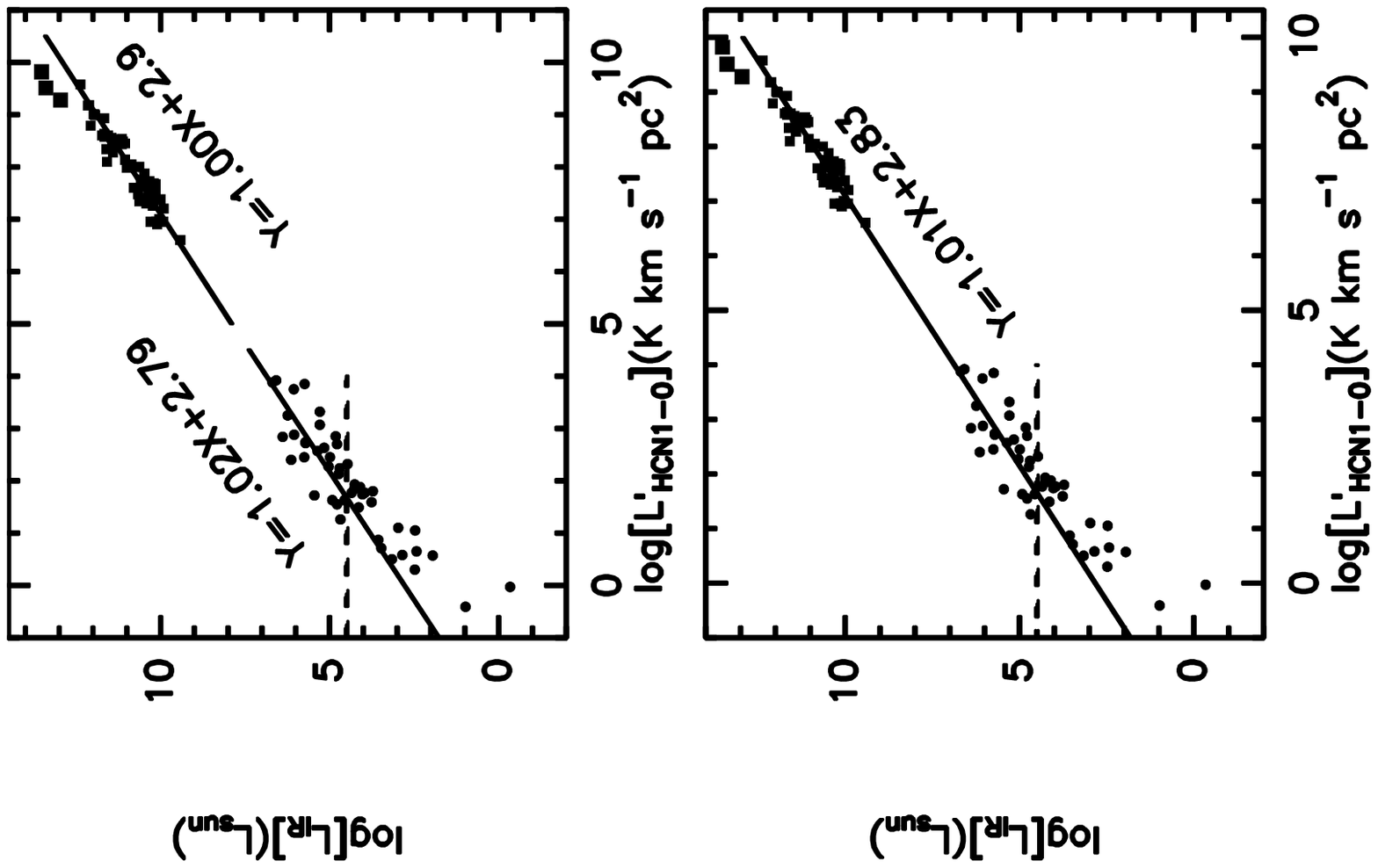}}
\caption{logL$_{IR}$ - logL$'_{HCN1-0}$ correlation for Galactic and
 extragalactic sources. Fig.1a (above) indicates the linear least squares 
 fit for Galactic cores (with $\lfir > 10^{4.5}$ L$_{\odot}$, points above 
 the dashed line) and for 
 galaxies, separately.  Fig. 1b (down) shows a overall fit for both parts.  
 The three isolated solid squares are high z HCN 1-0 points from 
 Solomon et al. (2003), Vanden Bout et al. (2004) and Carilli et al. (2005); 
 they are not included in the fit because the sources are 
 QSOs and the contribution from the AGN to \lfir\ is not yet clear. 
 }

\label{f1}
\end{figure}

\clearpage
\begin{figure}[hbt!]
\epsscale{0.75}
\rotatebox{-90}{\plotone{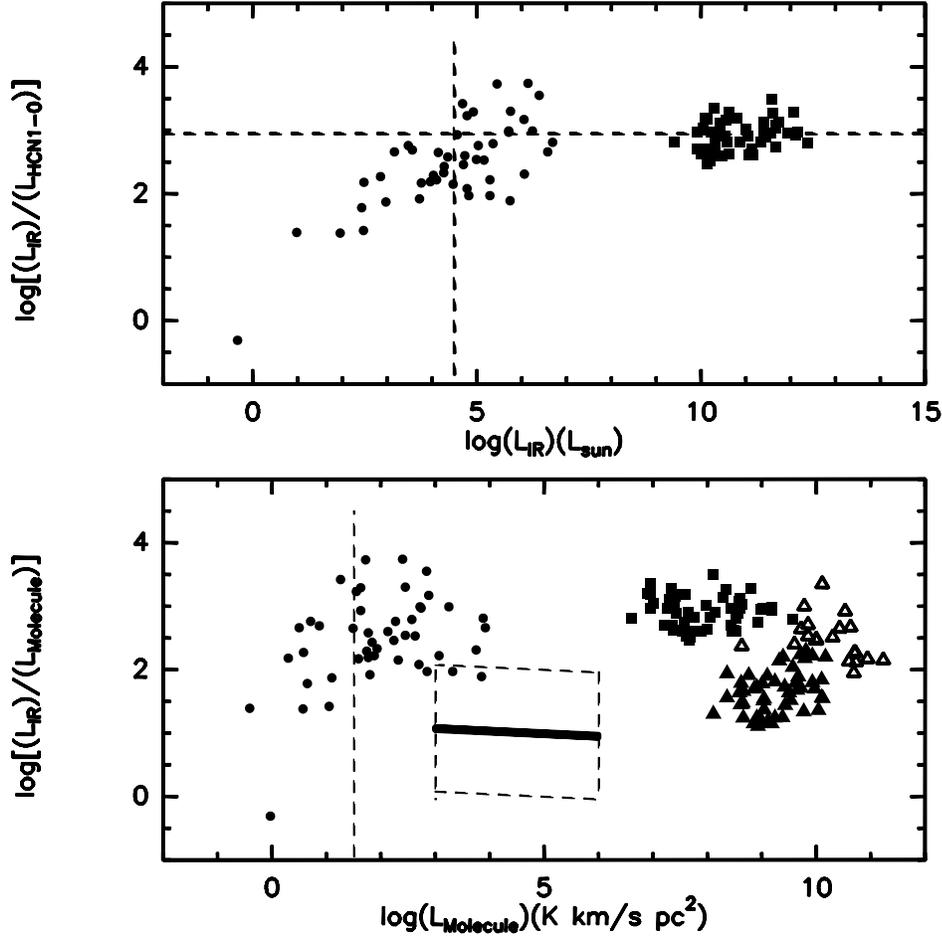}}
\caption{Above: Correlation between L$_{IR}$ and L$_{IR}$/L$_{HCN1-0}$ 
for galaxies (solid squares) and Galactic star forming cores (filled circles). 
 L$_{IR}$/L$_{HCN1-0}$ is constant for a large range of L$_{IR}$ utill 
 a turn off around L$_{IR}$=10$^{4.5}$ L$_{\odot}$.
Below: L$_{IR}$ per unit of molecular gas vs. molecular line luminosity. 
 The star formation rate per amount of CO gas changes a lot from Galactic GMCs
(the heavy line, with a dashed line boundary to indicate the variation), to
galaxies (filled triangles) and high-z CO galaxies (hollow 
triangles). }

\label{f2}
\end{figure}

\end{document}